 \documentclass[final,3p,times,twocolumn]{elsarticle}

\usepackage{graphicx}

\usepackage{amssymb}
\usepackage{wasysym}

\usepackage{lineno}

\journal{}

\begin{document}

\begin{frontmatter}



\title{Anisotropic diffusion of electrons in liquid xenon with application to improving the sensitivity of direct dark matter searches}
\author{Peter Sorensen}
\ead{pfs@llnl.gov}
\address{Lawrence Livermore National Laboratory, 7000 East Ave., Livermore, CA 94550, USA}
\begin{abstract}
Electron diffusion in a liquid xenon time projection chamber has recently been used to infer the $z$ coordinate of a particle interaction, from the width of the electron signal.  The goal of this technique is to reduce the background event rate by discriminating edge events from bulk events.  Analyses of dark matter search data which employ it would benefit from increased longitudinal electron diffusion.  We show that a significant increase is expected if the applied electric field is decreased.  This observation is trivial to implement but runs contrary to conventional wisdom and practice.  We also extract a first measurement of the longitudinal diffusion coefficient, and confirm the expectation that electron diffusion in liquid xenon is highly anisotropic under typical operating conditions.  
\end{abstract}

\end{frontmatter}

\begin{figure}[ht]
\centering
\includegraphics[width=0.48\textwidth]{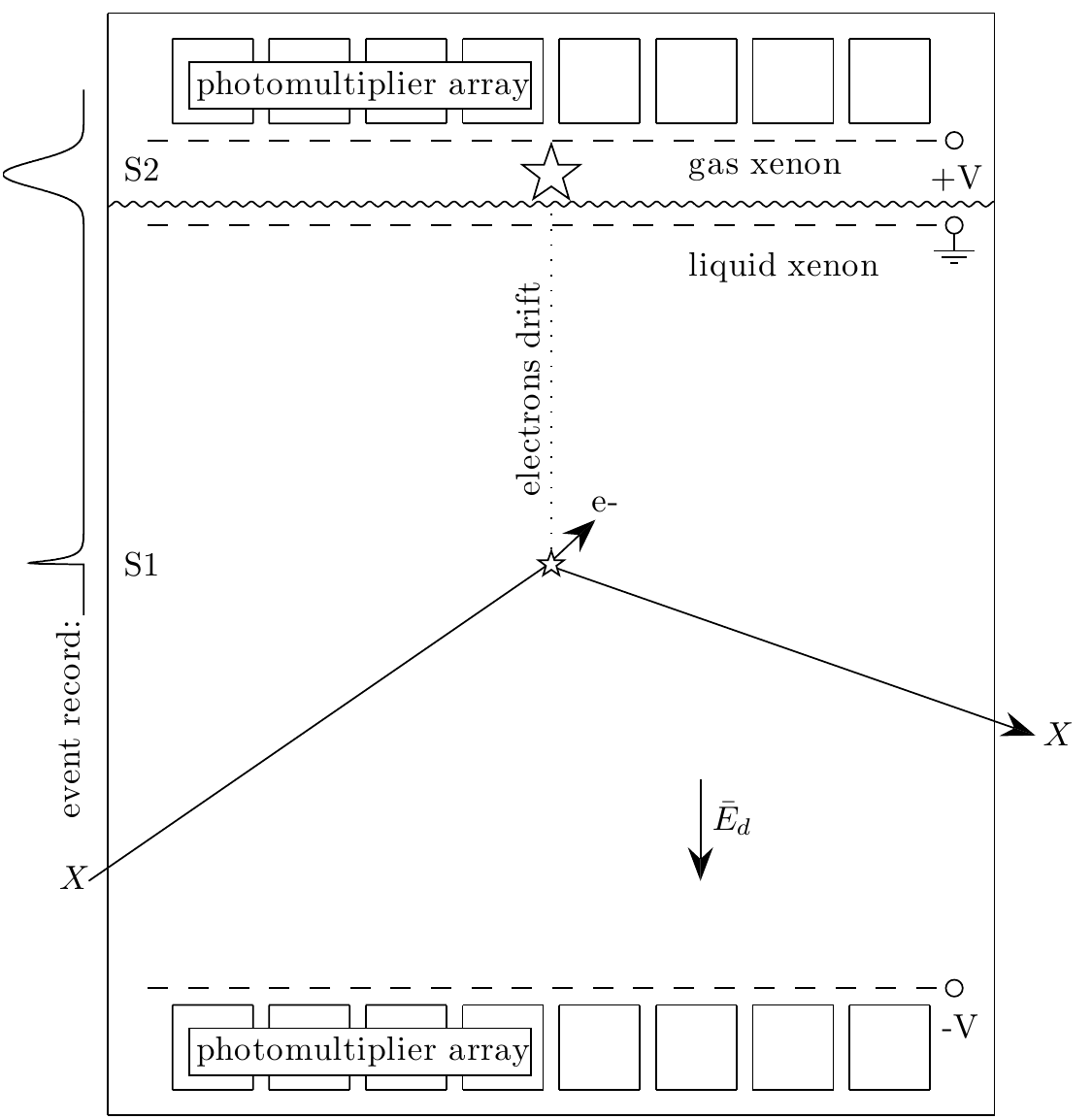}
\caption{Schematic representation of a liquid xenon time projection chamber, showing interaction of a particle $X$.  Primary scintillation photons (S1) are generated at the interaction vertex; electrons are also generated, and drifted to the gaseous xenon where they create proportional scintillation (S2).  A typical event record corresponding to such an interaction is shown along the left of the diagram. }
\label{fig0}
\end{figure}

The liquid xenon time projection chamber has proven to be an excellent detector for dark matter searches \cite{2010mckinsey,2010aprile}.  The principle of operation is illustrated in Fig. \ref{fig0}, and additional details can be found in {\it e.g.} \cite{2010aprile2}.  In standard analyses \cite{2010aprile,2009lebedenko,2009angle,2008angle, 2007alner}, the $z$ coordinate of interaction is obtained from the time delay between the electron (S2) and primary scintillation (S1) signals.  The delay results from the transit time of the electrons through the liquid.  The precision is $\sigma_{e}v_d\lesssim0.1$~cm, where $\sigma_{e}\sim0.3~\mu$s is the typical width of an S2 pulse, and the electron drift velocity is $v_d\sim0.2$~cm/$\mu$s \cite{1982gushchin}. S2 is measured via proportional scintillation. The $x-y$ coordinates of interaction are reconstructed from the hit pattern of S2 on the photomultipliers, with precision $\sim0.3$~cm \cite{2007angle}.  In this framework, electron diffusion has been safely ignored.  

Recent interest in the possibility that dark matter may be $\mathcal{O}(10)$~GeV has motivated the XENON10 collaboration to explore a new means of analyzing their dark matter search data \cite{2010sorensen}.  By ignoring S1 and requiring only S2 for dark matter candidate events, the detector retains full efficiency for nuclear recoils with energy as small as $\sim1$~keV.  This approach is prompted by the fact that the detection efficiency for S1 begins to fall below $\sim10$~keV \cite{2010sorensen2}.  In the absence of S1, the $z$ coordinate of interaction is indeterminate from the usual method, and incident particle type discrimination (from the ratio S2/S1) is lost.  However, the width of the S2 signal increases with drift distance through the liquid, due to the diffusion of a swarm of ionized electrons.  This allows an approximate $z$ coordinate to be assigned to each interaction, even in the absence of an S1 signal.  Rejection of high background event rate edge regions may therefore be partially recovered.  Interestingly, such S2-only analysis of dark matter search data finds a close parallel in the expected sub-keV signals from coherent neutrino-nucleus scattering.  Experiments which propose \cite{2004hagmann,2009akimov} to search for this signal may also benefit from $z$ coordinate reconstruction using electron diffusion.

The rate of transverse electron diffusion in liquid xenon is small, about 0.3~cm per meter of drift for an applied electric field $E_d\simeq1$~kV/cm \cite{1982doke}.  In this regime elastic electron collisions dominate, and solution of the Boltzmann equation predicts that diffusion is highly anisotropic \cite{1969skullerud}.  We are not aware of any measurement of the rate of longitudinal (i.e. parallel to $E_d$) electron diffusion.  In this brief article we show that the longitudinal diffusion coefficient $D_{L}$ of electrons in liquid xenon may be measured from existing data, and that electron diffusion in liquid xenon is highly anisotropic.  The precision in reconstructed $z$ coordinate using electron diffusion is $\sim$~cm, but improves with increasing $D_L$.  We show that a substantial increase in $D_L$ may be obtainable with order of magnitude reduction of $E_d$. Reducing the uncertainty in reconstructed $z$ coordinate would lead to increased efficiency of background rejection in S2-only light dark matter searches.  The importance of this is underscored by the loss of S2/S1 discrimination in such analyses.

\begin{figure}[ht]
\centering
\includegraphics[width=0.48\textwidth]{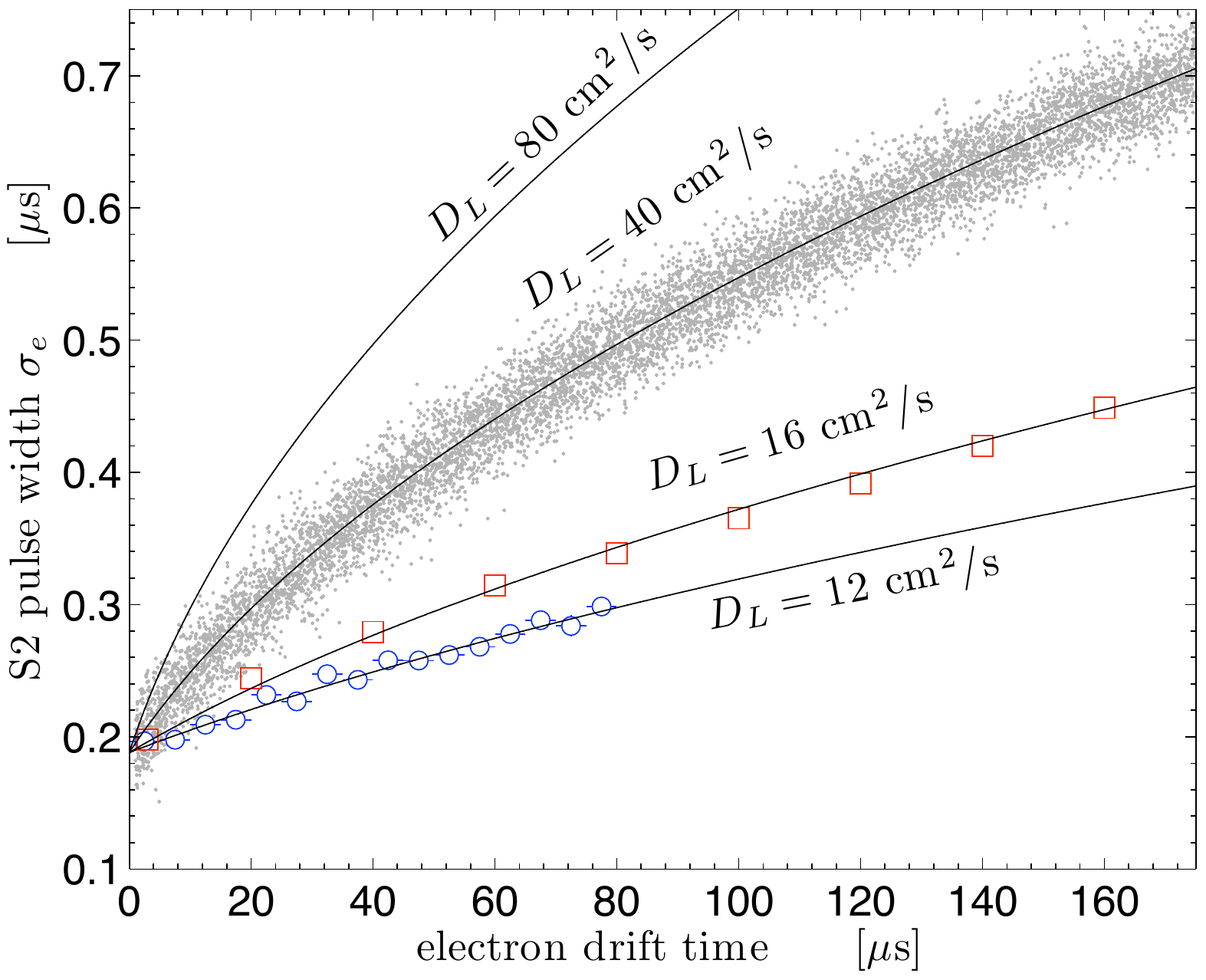}
\caption{Width of S2 pulses as a function of drift time through liquid xenon.  Measurements from XENON10 \cite{2010sorensen} at $E_d=730$~V/cm and $v_d=0.188$~cm/$\mu$s (circles, $1\sigma$ error smaller than data points) and XENON100 \cite{2010aprile3} at $E_d=530$~V/cm and $v_d=0.174$~cm/$\mu$s (squares).  The total drift distance was 15~cm and 30.6~cm, respectively.  $D_L$ was obtained from a fit of Eq. \ref{eq5} in each case.  The predicted width vs. drift time for two larger values of $D_L$ are shown, assuming $\sigma_0=0.19~\mu$s and $v_d=0.174$~cm/$\mu$s. A simulated distribution of events with Gaussian width $\sigma=0.02~\mu$s are shown for $D_L=40$~cm$^2$/s.}
\label{fig1}
\end{figure}

 Consider electrons drifting through liquid xenon in the $+z$ direction under the influence of an applied electric field $E_d$. S2 pulse width $\sigma_e$ from nuclear recoils is shown in Fig. \ref{fig1} as a function of drift time, for events in which an S1 signal was observed \cite{2010sorensen}.  The nuclear recoil energy range considered is $\mbox{E}_{nr}\lesssim30$~keV.  At these energies, the spatial extent of electronic excitation from a recoiling xenon nucleus is $<0.1~\mu$m \cite{2009dahl}, measured as {\it rms} radius of the track.  We therefore treat the initial source of $n_0$ electrons as point-like at $t=0$, and write their distribution (in co-moving coordinates) at a later time $t$ as \cite{1969parker,1974huxley}
\begin{equation} \label{eq1}
n = \frac{n_0}{\sqrt{4\pi D_{L} t}}\mbox{exp}(\frac{-z^2}{4 D_{L} t}).
\end{equation}
We have neglected the transverse dimensions, and note that a measurement of $D_{T}$ is given in \cite{1982doke}.  The mean square displacement of the electron distribution is
\begin{equation} \label{eq2}
\sqrt{\bar{z}^2} \equiv \sigma_z = \left( \frac{1}{n_0}\int_{-\infty}^{+ \infty} nz^2 dz \right) ^{1/2}
\end{equation}
from which one easily obtains
\begin{equation}\label{eq3}
\sigma_z = \sqrt{2 D_{L} t}.
\end{equation}
It is more natural to discuss $\sigma_t = \sigma_z/v_d$.  The measured $S2$ signal shape is a convolution of the Gaussian electron distribution defined by Eq. \ref{eq1}, and the average proportional scintillation distribution $f_0$ from a single electron.  Proportional scintillation is created in a $\sim0.3$~cm gap of gaseous xenon under high electric field \cite{2010aprile2}, with a roughly flat probability per unit path length.  We model $f_0$ as a boxcar function with standard deviation $\sigma_0$.  If $\sigma_t$ and $\sigma_0$ are similar to within $20\%$, a criteria that is satisfied for events in which the electrons have drifted for a time $t\gtrsim30~\mu$s, we can write 
\begin{equation}\label{eq4}
\sigma_e^2 \simeq \sigma_t^2 + \sigma_{0}^2
\end{equation}
for the measured width of the $S2$ signal.  Equation \ref{eq4} is exact if $f_0$ is Gaussian with standard deviation $\sigma_{0}$.  The diffusion coefficient $D_L$ can then be obtained from a fit to 
\begin{equation} \label{eq5}
\sigma_e = \sqrt{\frac{2 D_{L} t}{v_d^2}+\sigma_0^2},
\end{equation}
with $\sigma_0$ a free parameter.  We find $D_{L} = 12\pm1$~cm$^2$/s and $\sigma_0=0.190\pm0.005~\mu$s for the $E_d=730$~V/cm XENON10 data\footnote{The diffusion coefficient is often reported as $\widetilde{D}\equiv\sqrt{2D_L/v_d}$, which for the present case is about $34~\mu\mbox{m}~/\sqrt{\mbox{mm}}$}, fitting only the range $t\gtrsim40~\mu$s.  This is shown in Fig. \ref{fig1}, and the ratio $D_L/D_T=0.15$ is in good agreement with theoretical expectations \cite{1969skullerud}.  

\begin{figure}[ht]
\centering
\includegraphics[width=0.48\textwidth]{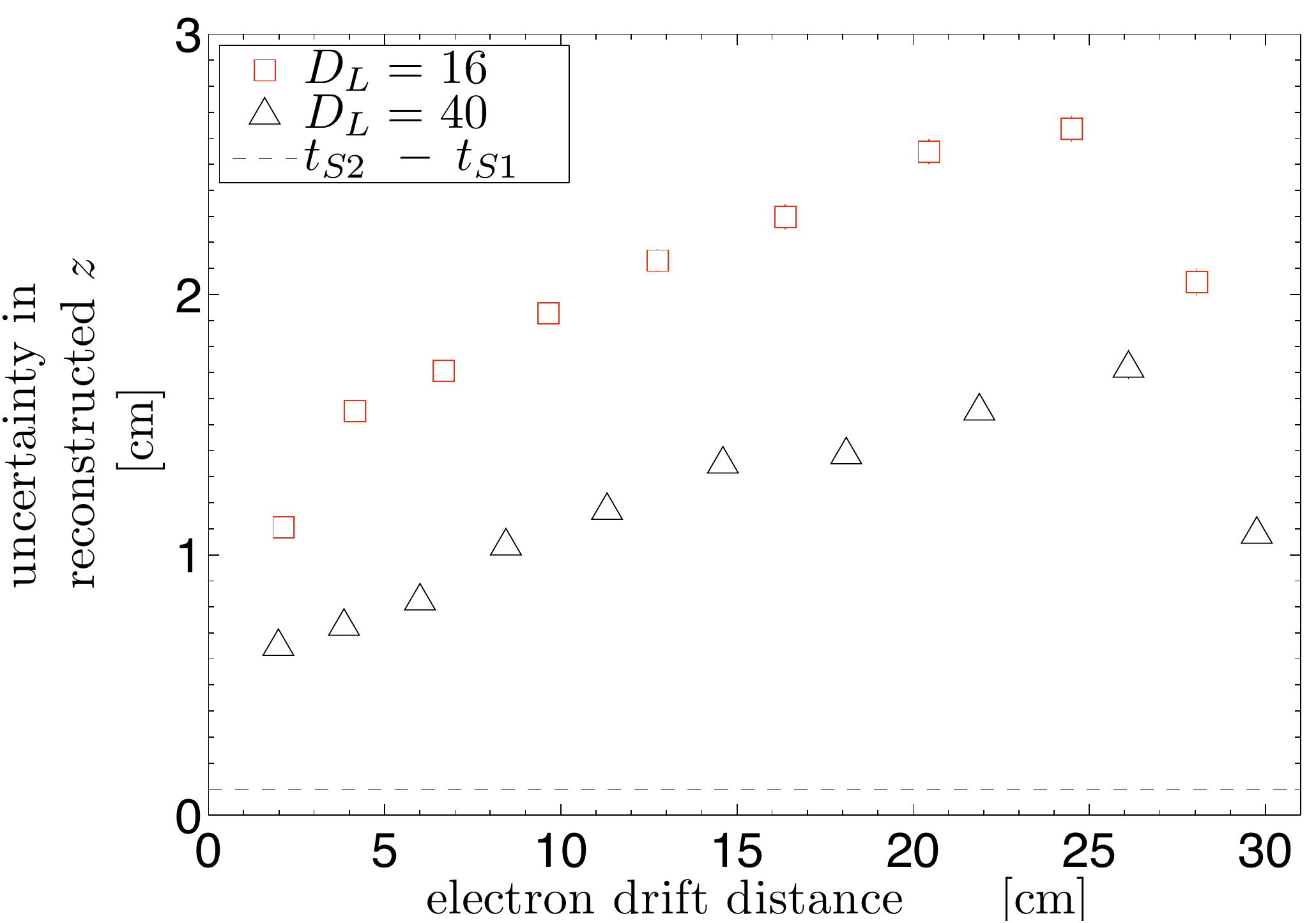}
\caption{$1\sigma$ uncertainty in $z$ coordinate reconstructed from S2 pulse width varies with $D_L$ as well as $z$.  Standard $z$ reconstruction from the delay time between S2 and S1 signals results in uncertainty $\sim0.1$~cm.}
\label{fig2}
\end{figure}

The electromagnetic background in liquid xenon dark matter detectors is presently dominated by radioactivity in the photomultipliers.  The primary utility of electron diffusion for light dark matter searches is thus to reject events at the top and bottom edges of the active region, near the photomultipliers. It can be seen from Fig. \ref{fig1} that in XENON10, a typical width $0.22~\mu$s  is measured for events which have drifted $20~\mu$s.  This corresponds to a depth of about 3.8~cm below the liquid surface.  However, the width $\sigma\simeq0.02~\mu$s of the distribution of $\sigma_e$ results in a $>1$~cm uncertainty in reconstructed $z$ coordinate, as shown in Fig. \ref{fig2}.  

 
In principle, improving the efficiency with which edge events are rejected in S2-only analyses could follow two paths:  decreasing the width of the $\sigma_e$ distribution, or increasing the electron diffusion.  The latter appears more accessible. The ratio $D_L/D_T$ approaches unity at very low electric field \cite{1969lowke}.  We therefore expect that as $E_d$ is decreased, $D_L$ will approach the $\sim80$~cm$^2$/s low-field limit of $D_T$ \cite{1982doke}.  In Fig. \ref{fig1} we show preliminary electron diffusion data obtained by the XENON100 experiment (stars, uncertainty not shown) at $E_d=530$~V/cm \cite{2010aprile3}.  As expected, this data shows an increase in $D_L$.

Dark matter searches with liquid xenon detectors have tended to operate at the highest possible value of $E_d$, limited by the maximum voltage that could be applied to the cathode.  The motivation for this is the observation \cite{2009lebedenko} of improved S2/S1 discrimination power at higher $E_d$, although other work has found no significant improvement \cite{2009dahl}.  For light dark matter searches using only the S2 signal, it would be beneficial to operate at the lowest practical value of $E_d$.  An optimal value of $E_d$ will maximize $D_L$ without substantially reducing the ionization yield.  Studies of the ionization yield of liquid xenon for nuclear recoils indicate only a 10\% decrease if $E_d$ is reduced from 2~kV/cm to 100~V/cm \cite{2006aprile}.  The diffusion coefficient at this low drift field is presently unknown.

\begin{figure}[ht]
\centering
\includegraphics[width=0.48\textwidth]{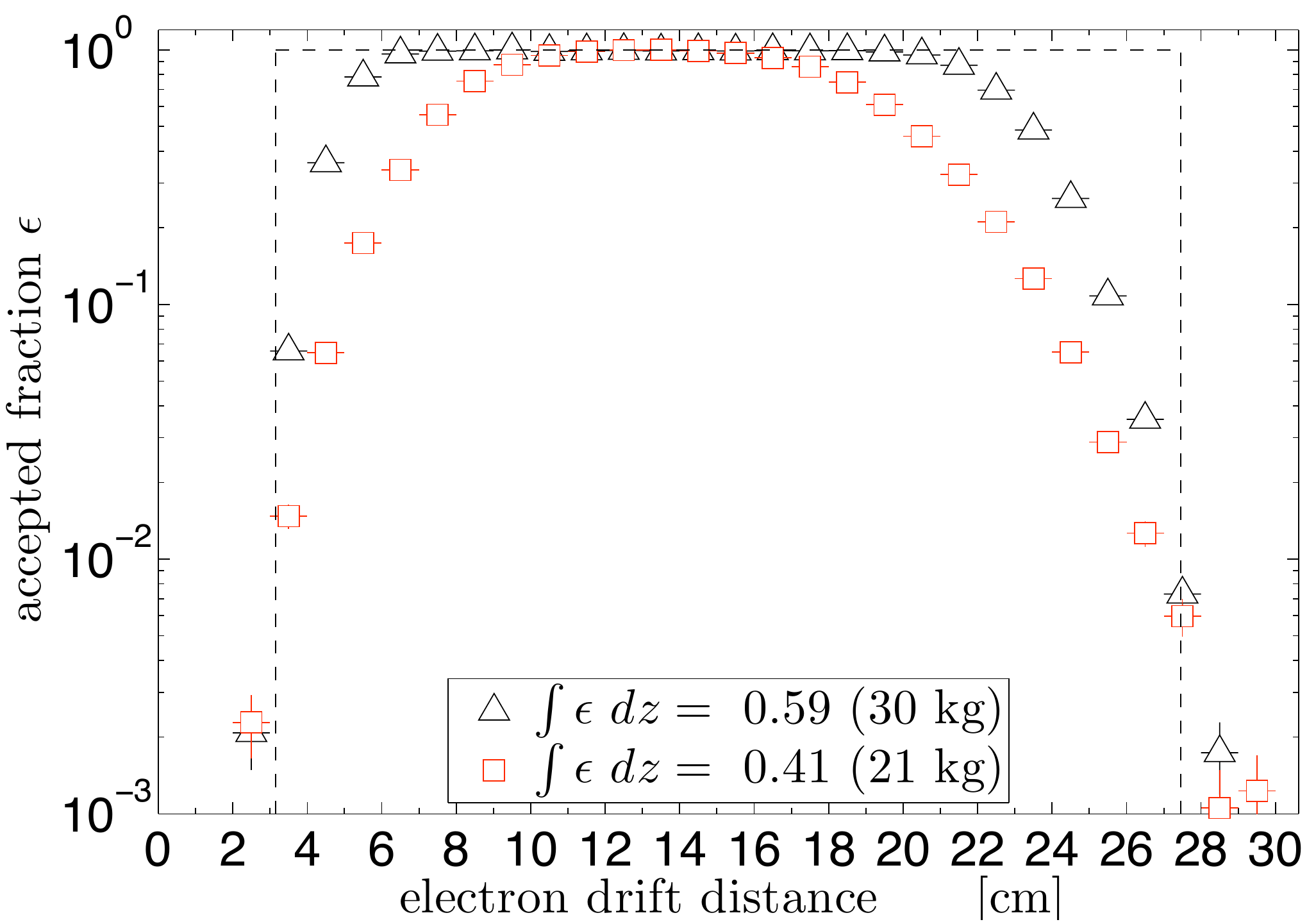}
\caption{Predicted fraction of accepted events in XENON100, with $z$ coordinate reconstructed from S2 pulse width, for $D_L=16$~cm$^2$/s ($\Box$) and $D_L=40$~cm$^2$/s ($\triangle$).  $\sigma_e$ cut bounds for each case are given in the text.  Dashed lines indicate the accepted range of $z$ coordinate in \cite{2010aprile}, using standard $z$ reconstruction.}
\label{fig3}
\end{figure}

Suppose $D_L$ were as high as 40~cm$^2$/s at 100~V/cm.  A simulated distribution of events with width $\sigma=0.02~\mu$s in shown in Fig. \ref{fig1}. In this case, events within $\sim3$~cm of the edges of the XENON100 detector would be rejected at $3\sigma$ by requiring $\sigma_e>0.31~\mu$s and $\sigma_e<0.61~\mu$s.  The $z$-dependent efficiency of this cut is shown in Fig. \ref{fig3} ($\triangle$), and results in an effective target mass of 30~kg.  This may be compared with the 40~kg target mass obtained by the XENON100 Collaboration from standard $t_{S2}-t_{S1}$ $z$ coordinate reconstruction (Fig. \ref{fig3}, dashed).  Achieving the same level of background reduction with $D_L=16$~cm$^2$/s would require $\sigma_e>0.27~\mu$s and $\sigma_e<0.39~\mu$s, and the effective target mass would be only 21~kg (Fig. \ref{fig3}, $\Box$).  

The sensitivity of liquid xenon detectors to light dark matter using the S2-only analysis technique will ultimately depend on both their ability to reject edge events, on the ionization yield per keV of nuclear recoil energy \cite{2010sorensen} and on the S2 signal threshold.  The latter is presently reported to be about 15 electrons \cite{2010aprile}, significantly above the single electron threshold obtained by XENON10 \cite{2010sorensen}.  But while XENON10 was decommissioned in 2007 with all of its data acquired at $E_d=730$~V/cm, XENON100 has the potential to improve an important aspect of their sensitivity by turning a single knob.  With a sufficiently low background event rate in the central target, liquid xenon detectors may have the potential for unambiguous detection of light dark matter.


\section*{Acknowledgments}
I am grateful to Adam Bernstein for a critical reading of the manuscript, to Mike Heffner for interesting discussions about diffusion, and to the reviewer for helpful suggestions.





\bibliographystyle{elsarticle-num}
\bibliography{<your-bib-database>}

\begin{thebibliography}{00}

\bibitem{2010mckinsey} D.N. McKinsey {\it et al.} (LUX Collaboration), J. Phys.: Conf. Ser. 203 (2010) 012026.
\bibitem{2010aprile} E. Aprile {\it et al.} (XENON100 Collaboration), Phys. Rev. Lett. {\bf105} 131302 (2010).
\bibitem{2009lebedenko} V.N. Lebedenko {\it et al.} (ZEPLIN III Collaboration), Phys. Rev. D {\bf80} 052010 (2009).
\bibitem{2009angle} J. Angle {\it et al.} (XENON10 Collaboration), Phys. Rev. D {\bf80}, 115005 (2009).
\bibitem{2008angle} J. Angle {\it et al.} (XENON10 Collaboration), {Phys. Rev. Lett.} {\bf101} 091301 (2008).

\bibitem{2007alner} G.J. Alner {\it et al.} (ZEPLIN II Collaboration), Astropart. Phys. {\bf28} 287 (2007).
\bibitem{1982gushchin} E.M. Gushchin, A.A. Kruglov and I.M. Obodovskii, Sov. Phys. JETP {\bf4} 55 (1982).
\bibitem{2007angle} J. Angle {\it et al.} (XENON10 Collaboration), Nucl. Phys. B (Proc. Suppl.) {\bf173} 117 (2007).
\bibitem{2010sorensen} P. Sorensen {\it et al.} (XENON10 Collaboration), PoS (IDM2010) 017, arxiv:1011.6439 
\bibitem{2010sorensen2} P. Sorensen, JCAP {\bf09} 033 (2010).

\bibitem{2004hagmann} C. Hagmann and A. Bernstein, IEEE Trans. Nucl. Sci. {\bf51} 2151 (2004).
\bibitem{2009akimov} D. Akimov, A.~Bondar, A.~Burenkov and A.~Buzulutskov, JINST {\bf4} P06010 (2009).
\bibitem{1982doke} T. Doke, Nucl. Instr. and Meth A {\bf196} 87 (1982).
\bibitem{1969skullerud} H.R. Skullerud, J. Phys. B (Atom. Molec. Phys.) {\bf2} 696 (1969). 
\bibitem{2009dahl} C.E. Dahl, Ph.D. Thesis (2009), Princeton University, Princeton NJ. 

\bibitem{1969parker} J.H. Parker, Jr and J.J. Lowke, Phys. Rev. {\bf181} 290 (1969).
\bibitem{1974huxley} L.G.H. Huxley and R.W. Crompton, {\it The Diffusion and Drift of Electrons in Gases} (Wiley, New York, 1974).
\bibitem{2010aprile2} E. Aprile  {\it et al.} (XENON10 Collaboration), arxiv:1001.2834 
\bibitem{1969lowke} J.J. Lowke and J.H. Parker, Jr, Phys. Rev. {\bf181} 302 (1969).
\bibitem{2010aprile3} E. Aprile, seminar at WONDER2010, http://wonder.lngs.infn.it

\bibitem{2006aprile} E. Aprile {\it et al.}, Phys. Rev. Lett. {\bf97} 081302 (2006).

\end{thebibliography}



\end{document}